# Deposition and patterning of magnetic atom trap lattices in FePt films with periods down to 200 nm

A. L. La Rooij, S. Couet, M. C. van der Krogt, A. Vantomme, K. Temst, and R. J. C. Spreeuw

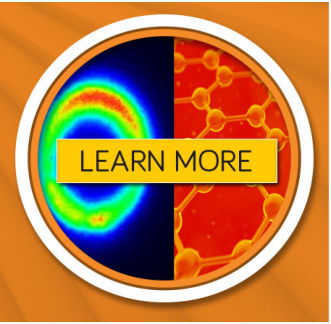

# Deposition and patterning of magnetic atom trap lattices in FePt films with periods down to 200 nm

A. L. La Rooij,[1,a)] S. Couet,[2] M. C. van der Krogt,[3] A. Vantomme,[4] K. Temst,[4] and R. J. C. Spreeuw[1,b)]

[1]*Van der Waals - Zeeman Institute, Institute of Physics, University of Amsterdam, Science Park 904, 1098 XH Amsterdam, The Netherlands*
[2]*Imec, Leuven, Belgium*
[3]*Kavli Nanolab Delft, Quantum Nanoscience, TU Delft, Delft, The Netherlands*
[4]*Instituut voor Kern-en Stralingsfysica, KU Leuven, Leuven, Belgium*

We report on the epitaxial growth and the characterization of thin FePt films and the subsequent patterning of magnetic lattice structures. These structures can be used to trap ultracold atoms for quantum simulation experiments. We use molecular beam epitaxy to deposit monocrystalline FePt films with a thickness of 50 nm. The films are characterized with X-ray scattering and Mössbauer spectroscopy to determine the long range order parameter and the hard magnetic axes. A high monocrystalline fraction was measured as well as a strong remanent magnetization of $M = 900$ kA/m and coercivity of 0.4 T. Using electron beam lithography and argon ion milling, we create lattice patterns with a period down to 200 nm, and a resolution of 30 nm. The resulting lattices are imaged in a scanning electron microscope in the cross-section created by a focused ion beam. A lattice with continuously varying lattice constant ranging from 5 $\mu$m down to 250 nm has been created to show the wide range of length scales that can now be created with this technique. © *2018 Author(s). All article content, except where otherwise noted, is licensed under a Creative Commons Attribution (CC BY) license (http://creativecommons.org/licenses/by/4.0/).* https://doi.org/10.1063/1.5038165

## I. INTRODUCTION

Lattices of trapped, cold atoms play a central role in the development of quantum simulators and quantum information protocols. Implementations on the scale of several micrometers typically employ long-range interactions of Rydberg atoms,[1,2] while those on the submicrometer scale are described by Hubbard models, governed by tunneling and on-site interaction.[3] Based on the trapped neutral atoms, quantum simulators of spin models in many lattices and low dimensional systems have been successfully realized.[4] While optical lattices have so far been the main method of choice, the lattices of magnetic micro- or nanotraps on an atom chip provide promising alternatives and offer additional possibilities. Magnetic lattices can be created over a large range of length scales and provide virtually unlimited freedom in designing 2D lattice geometries.

Standard optical lattices have periods of one half wavelength of the lattice laser, typically 400 nm. There is a strong interest to scale down these atomic trapping lattices, in particular, for experiments simulating Hubbard and related models.[5,6] With smaller inter-trap spacing or lattice constants, higher tunneling rates and stronger interactions over multiple sites may be achieved, which will allow the study of more complicated models such as the extended Hubbard model and long range spin models.[7,8]

Proposals for downscaling include nanoplasmonic systems, arrays of superconducting vortices, and optical lattices with additional dressing fields.[9] Magnetic lattices form a natural candidate to generate lattices at these subwavelength scales.[10,11] Scaling down is only limited by the resolution of the lithographic process by which they are made. As we show here, these nanofabricated lattices can even be scaled down continuously, in the form of a tapered lattice where the lattice constant varies across the atom chip. Furthermore, with the recently developed design methods, novel trapping geometries can be created.[12]

This paper focuses on the fabrication of small magnetic lattices to create atom traps with the highest possible resolution, provided by electron beam lithography (EBL). In the previous experiments, magnetic lattices with the periods of 10 $\mu$m or larger were used both in one and in two dimensions to trap ultracold gases.[13–15] These experiments used permanent magnetic films with a thickness of several hundreds of nanometers, patterned by optical lithography. For patterning at the sub-micron scale, EBL can be used. This technique has been well developed for semiconductor applications based on Si but has not been used so far for the fabrication of patterns in FePt. In the Swinburne group, a magnetic lattice with a period of 700 nm was created this way in a Co/Pd multilayer film of 10.3 nm thick.[16]

We first describe the fabrication of 50 nm thick monocrystalline magnetic films of FePt by molecular beam epitaxy (MBE), and their characterization. We then describe the patterning process, to create lattices with periods down to 200 nm, well within the regime of sub-wavelength tunneling physics. A new process was developed to fabricate FePt structures with a resolution of 30 nm. With the patterning, we created various lattice geometries over a range of length scales, including square lattices with lattice constants 900,

[a)]a.l.larooij@uva.nl
[b)]r.j.c.spreeuw@uva.nl

600, and 200 nm, and a Kagome lattice with 5 $\mu$m lattice constant. We also fabricated the tapered structures spanning all length scales at which magnetic lattices have been made.

This paper is divided into three different sections. In Sec. II, we describe the newly developed FePt films and their characterization. In Sec. III, we describe the lithographic patterning process for the trapping arrays in the FePt film, and in Sec. IV, we discuss the resulting magnetic patterns.

## II. MOLECULAR BEAM EPITAXY OF MONO-CRYSTALLINE FePt FILMS

To create traps for ultracold atoms, one needs to carefully consider the magnetic properties of the material. The coercivity must be higher than the externally applied magnetic fields of up to 0.02 T. The magnetization hysteresis curve should ideally be as square as possible. The remanent magnetization should withstand the bake-out procedure that is required to obtain ultra high vacuum (UHV) conditions, typically at 150 °C. The choice for FePt as a hard magnetic material is based on the experience in the previous experiments. The alloy FePt in the $L1_0$ crystalline phase has a high remanent magnetization and coercivity, both above 0.1 T depending on the fabrication parameters.[17–19]

Previously, the films of FePt ranging in thickness from 200 to 400 nm were deposited by sputtering techniques.[20] These films were not compatible with the structures in the few 100 nm range, due to their thickness, flatness, and grain size. Therefore, new films were deposited by molecular beam epitaxy (MBE). This technique was explored by various research groups[17,21–23] to develop magnetic recording media from FePt and is also used to study the interfaces in magnetic materials.[24] The slower MBE deposition process in combination with a substrate of which the atomic crystal spacing is epitaxially matched helps to create higher quality magnetic films. Our aim is to fabricate the microtraps with lattice spacings ranging from several micrometers down to 200 nm. Films of 50 nm thick were fabricated to facilitate the atomic traps at all the above-mentioned length scales. More details on the resulting trap parameters can be found in Ref. 25.

### A. MBE deposition of FePt

To synthesize mono-crystalline films, a matched substrate of MgO was used to create the FePt crystal into the desired (100) orientation. In particular, thin substrates of only 150 $\mu$m thick MgO were used to establish a thin chip for future atomic physics experiments. Polished MgO substrates were used with a maximum roughness of less than 2 nm/$\mu$m, measured with an interferometer microscope.[26]

The substrates of $20 \times 20$ mm$^2$ were pre-annealed overnight in a vacuum oven that was connected to the MBE chamber. The samples were heated to 500 °C to allow direct growth into the $L1_0$ phase of FePt. By controlling the growth rate around 0.5 Å/s, we obtained mono-crystalline FePt films with the same roughness as the original substrate. Six batches of two substrates were made, all with an equal ratio of Fe and Pt to create Fe$_{50}$Pt$_{50}$. The thickness of the films was controlled by timing the growth rate *in situ*, and all films were found to be 50(2) nm thick.

### B. Characterization of FePt films

The quality of the films was investigated by several methods. X-ray diffraction was performed on all samples to determine the long range crystal and chemical order of the FePt crystals. We define a long range order parameter $S = \sqrt{0.45 \times I_{100}/I_{002}}$.[19] Here, $I_{100}$ and $I_{002}$ are the intensities of the X-ray peaks corresponding to the (100) direction and the (002) direction. A perfectly ordered mono-crystal will give $S = 1$. We found S-values of 0.63, 0.55, and 0.63 for batches 1, 2, and 3, respectively, indicating a high mono-crystalline fraction. The chip that was fabricated to host the first nanolattices was made out of batch 3, the diffractogram of which is shown in Fig. 1.

We also performed Mössbauer spectroscopy, the results of which are presented in Fig. 2. From the integral of the peaks in this spectrum, one can find the orientation of the magnetic axes. The angle between the main crystal axes (100) and the magnetic orientation $\theta$ can be fitted to the ratio of the intensity of the Mössbauer peaks by the following formula, from Ref. 27:

$$3 : \frac{4\sin^2\theta}{1+\cos^2\theta} : 1 : 1 : \frac{4\sin^2\theta}{1+\cos^2\theta} : 3. \quad (1)$$

For example, the intensity ratio for a $^{57}$Fe crystal which is magnetized in-plane ($\theta = 90°$) is 3:4:1:1:4:3. In the fit to the spectrum of the deposited mono-crystalline FePt samples, we see small signs of a 2nd and 5th peaks. We attribute this to a predominantly out-of-plane magnetization, consistent with the large but not perfect long range order parameter (see Fig. 2). These measurements match the results of previously grown samples.[19]

Additional characterization of the magnetic properties of the films was performed using SQUID measurements at both 300 K and 350 K, without any observable difference. The 300 K hysteresis curve is shown in Fig. 3. The curve has close to rectangular shape, indicating both a high remanent

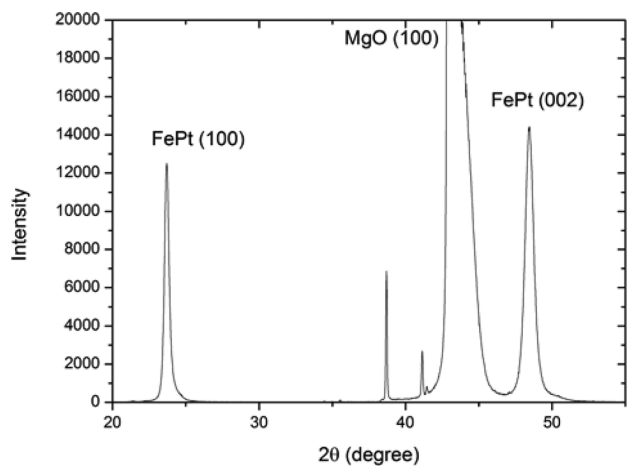

FIG. 1. X-ray diffraction of the MBE-grown FePt films.

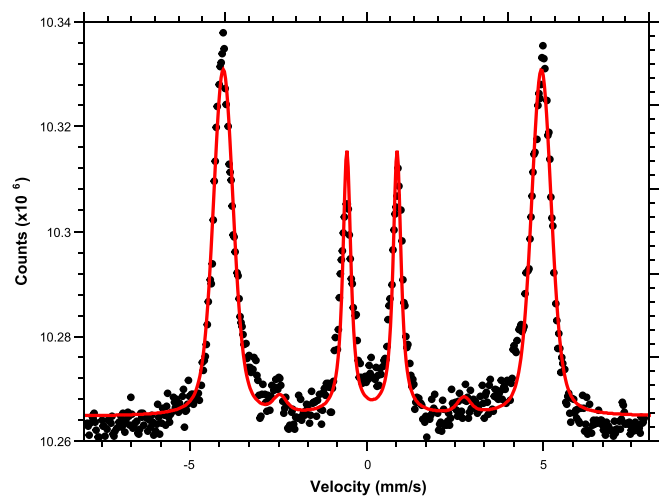

FIG. 2. Mössbauer spectrum of the FePt film from batch 3, measured at KU Leuven. The red line is a fit with the RECOIL software.[28]

magnetization of $M = 900 \, \text{kA/m} = 900 \, \text{emu/cm}^3$ and a high coercivity of 0.4 T.

Bake-out tests were performed in Amsterdam, by measuring hysteresis curves using the Magneto-Optical Kerr Effect (MOKE), both before and after heating the samples to 150 °C. No loss of the magnetization could be observed from these MOKE measurements. In the previous experiments, this loss was estimated to be around 3%.[29] The MOKE data were consistent with the SQUID hysteresis curves.

The grain size was measured with a scanning electron microscope (SEM) inspection of the samples. For comparison in Fig. 4, we show an SEM image of the MBE grown material together with that of an older sputtered FePt film that has been used previously for magnetic atoms chips.[30] In the sputtered material, grainy structures with a typical size of 40 nm can be observed. Even at a much higher resolution, a similar grain pattern is not visible on the MBE grown samples. A comparison between the parameters of the newly MBE grown films and older sputtered films is presented in Table I.

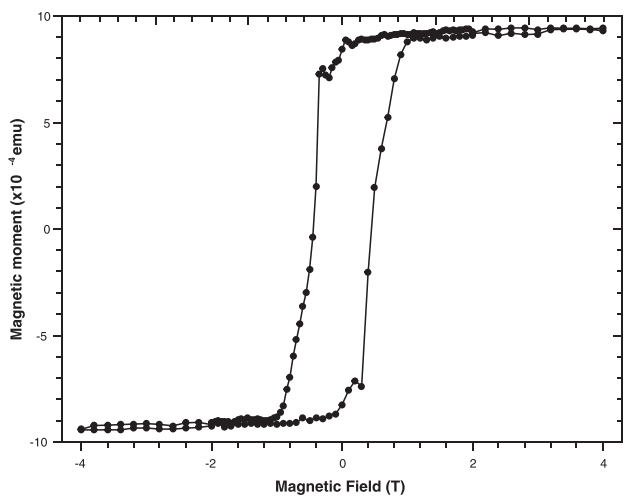

FIG. 3. Hysteresis curve of the FePt film from batch 3 at a temperature of 300 K, measured using a SQUID.

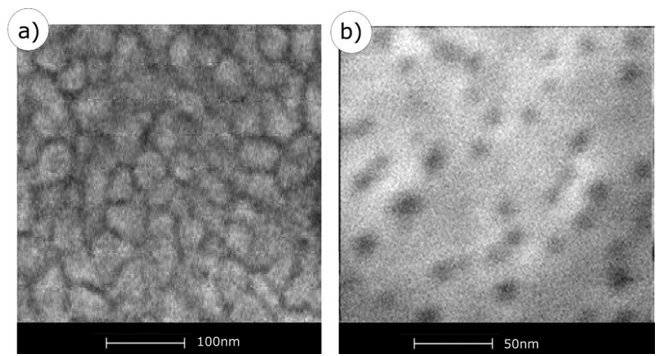

FIG. 4. SEM images of FePt samples. (a) Sputtered FePt with grains of approximately 40 nm. (b) Mono-crystalline MBE-grown FePt without any grain structure on this scale.

In summary of this section, we deposited an FePt film on the polished MgO substrates to fabricate nanoscale magnets using lithography. With X-ray diffraction, we found the magnetic films to have a long range order parameter of $S > 0.55$ out of 1 indicating a large mono-crystalline fraction. This was confirmed when we inspected the samples in an SEM. With Mössbauer spectroscopy, we found that the magnetic orientation angle was close to 0 indicating an out-of-plane magnetization. With a SQUID, we measured a square hysteresis curve that is required to create stable atomic traps. We found a strong magnetization as well as a large coercivity. By using a MOKE setup, we looked for the loss of magnetization after heating the samples but no such effect was found. Combined, these measurements indicated that the films had the excellent conditions to be etched into small nanoscale magnetic structures.

## III. PATTERNING PROCESS

### A. Electron beam lithography

In order to achieve the desired high resolution, we defined the lithographic patterns using electron beam lithography (EBL) at the Kavli Nanolab, Delft. The resolution needed to write lattices down to a few 100 nm is a few 10 nm, which is achievable using electron beams down to a few nanometers in size and a suitable electron-sensitive resist film. Initial attempts were made with hydrogen silsesquioxane (HSQ), because this resist should provide the highest resolution. However, on the thin monocrystalline MgO-FePt samples two complications occurred. First, the dose required for ideal HSQ development fluctuated strongly. We suspect this is due to the chemical instability of the HSQ solution, which is highly sensitive to temperature fluctuations.[31] Second, adhesion problems resulted in partial detachment from the sample during development. If the future experiments require the resolution or sensitivity of HSQ, the adhesion to the surface needs to be optimized to ensure a reproducible and stable development.

TABLE I. Properties of different FePt films that have been used to fabricate the magnetic lattices. The first two columns concern FePt films that have been sputtered at the University of Amsterdam[18] and at Hitachi.[30] The last column lists the properties of the MBE grown films that were fabricated in Leuven.[19]

| Property | UvA Sputtered | Hitachi Sputtered | Leuven MBE |
|---|---|---|---|
| $M_r$ (kA/m), remanent | 580 | 670 | 900 |
| $M_s$ (kA/m), saturation | 725 | 720 | 935 |
| $M_r/M_s$ | 0.80 | 0.93 | 0.95 |
| $H_c$ (G), coercivity | 9500 | 9500 | 4000 |
| $M_r$ decrease after 3 h bake at 150 °C | 20% | 3% | $\leq 1\%$ |
| Estimated grain size (nm) | 20 | 35 | … |
| Surface roughness (nm) | 1 | 6 | 2 |
| Thickness (nm) | 300 | 250 | 50 |

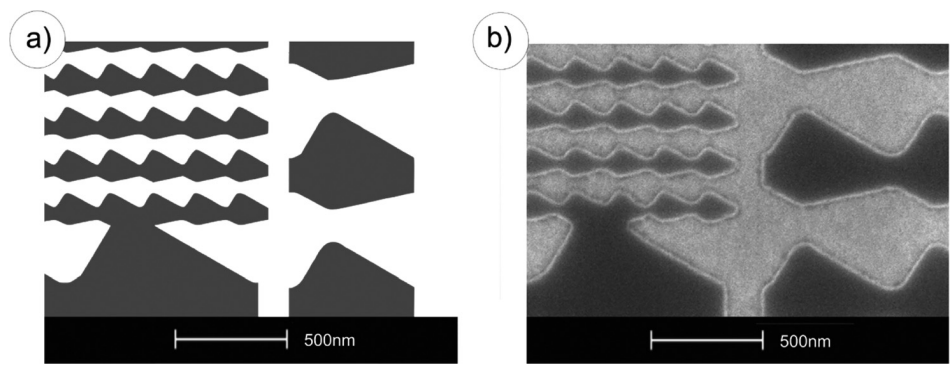

FIG. 5. (a) Pattern sent to the EBPG to write square lattices on three different length scales. In this view, segments with a period of 900, 600, and 200 nm are visible. (b) The developed pattern in the resist using the optimal dose of 1375 $\mu$C/cm$^2$. The scale bars are 500 nm.

For these reasons, we switched to AR-N.7500.08, a more stable and less sensitive resist.[32] This negative resist allows a resolution of approximately 30 nm. The recipe we used to create a 100 nm thick etch mask with AR-N.7500.08 at its ultimate resolution is as follows.

1. Spin a film of adhesion promotor.
2. Bake for 2 min at 200 °C.
3. Spin the AR-N.7500.08 at 4000 rpm to get a 100 nm thick film.
4. Bake for 1 min at 85 °C.
5. Expose at a dose of 1375 $\mu$C/cm$^2$.
6. Develop for 1 min in pure MF322.
7. Rinse for 1 min in 1 : 9 solution of MF322:H$_2$O.

An optimal dose for this negative resist was found at 1375 $\mu$C/cm$^2$, which produced structures with a resolution of 30 nm, while using a beam step size of 3 nm and a 5 nm spot. With this resolution, lattices were created down to 200 nm period as shown in Fig. 5. In Fig. 5(b), three square lattices are written with lattice spacings of 900, 600, and 200 nm. The lattices have a minimum open trench of 50 nm in the smallest 200 nm lattice. In lattices with periods below 200 nm, the resist structures connect because the width of the trenches becomes of the order of the resolution.

### B. Ultra low pressure plasma etching

The resist mask is transferred to the magnetic film by ion plasma etching. The selectivity of the etch rate for the magnetic layer as compared to the mask is an important parameter. The mask should ideally be just thick enough since some of the removed FePt falls back onto the sample during etching and builds up against the side of the left-over resist. The crystalline phase of this redeposited FePt, and therefore the magnetic behavior, is not known. Another mechanism inherent to perpendicular ion etching is that the sidewalls become sloped (facet formation). The sidewall slope $\alpha$ on the structure side is 30° for optimal plasma parameters. With this slope and the height of the magnetic layer of 50 nm, this etching technique is limited to the fabrication of lattices with a spacing of >200 nm (see Fig. 6). A method to achieve smaller edge slopes is to use a non-focused ion beam (FIB) instead of a plasma, which can etch under an angle with the sample. By varying the angle of incidence of the ion beam with a rotating sample, the structures with the straight or under etched (negative) slopes can be created.

An inductively coupled plasma (ICP) etcher (Surface Technology Systems) was used at a pressure of 1.6 mTorr. At these low pressures, we were able to obtain the sidewall slopes of 30°. The following etch settings were used: exposure of 210 s to a plasma with an argon flow of 35 SCCM. The ions were accelerated by a bias voltage of 260 V resulting from an ICP source power of 795 W and a platen power of 27.9 W.

To limit overetching, a laser interferometer was used to monitor the etching process and the process was stopped several seconds after the MgO substrate became visible. The inspection of cross sections of the samples showed no significant overetching. The sample shown in Fig. 7 was etched for 5 min to indicate the sidewall slopes more clearly. Also, the redeposition is visible as peaks that stick out above the magnetic layers, built up against the remaining resist. The final chip patterns and their cross sections are presented in Sec. IV.

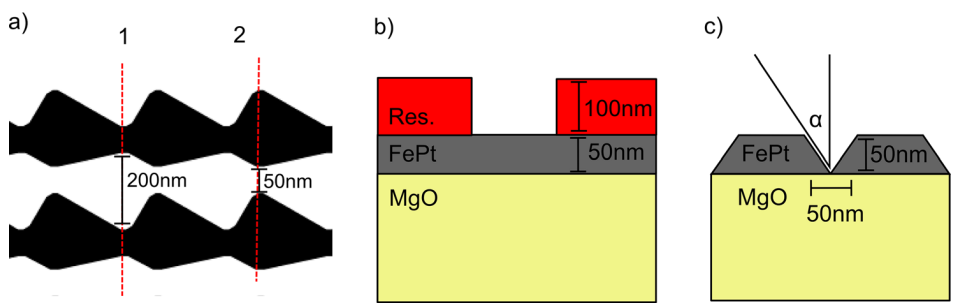

FIG. 6. (a) Pattern for a square lattice. Lines 1 and 2 indicate the positions of smallest trench and smallest standing resist structure. (b) Cross section of the pattern at position 1 before etching. (c) Etch slopes as they are created by plasma etching with optimal redeposition angle $\alpha = 30°$ at position 2.

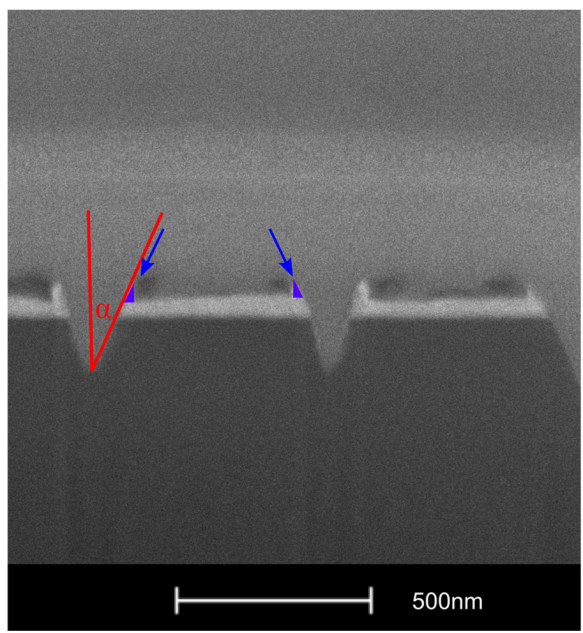

FIG. 7. A SEM image of a cross section made by focused ion beam milling. This is a 600 nm lattice as etched by the procedure described in the text. From top to bottom: Pt deposited locally to protect the interface during milling (grey), resist leftovers (dark), FePt (light), and MgO substrate (black). The etch slope is indicated in red with an angle of 30°. The overetched trenches go up to 200 nm into the MgO substrate. On the middle magnet, the redeposition against the resist is colored blue and indicated by the two arrows.

The redeposited FePt could not be removed chemically. The leftover resist and the redeposited FePt remained on the sample. These unwanted structures in the profile of the chip are hard to remove. To partially remove the redeposited FePt, one can use the chemically enhanced etching to assist the removal of sputtered Fe and Pt particles, as was recently investigated.[22] The final fabrication step is to evaporate a 50 nm thick Pt film on the sample surface to protect the FePt structures and to reflect optical beams in the experiments (mirror layer). The influence of this layer on trapped atoms close to the surface is discussed in detail in Refs. 10, 33, and 34. The inspection, which is presented in Sec. IV, was carried out before the samples were covered by this layer.

## IV. RESULTS

### A. Focused ion beam inspection

To inspect the fabricated structures, cross sections were made by focused ion beam (FIB) milling. Before these cross sections were made, Pt was deposited locally to protect the interface of the top layer. The Pt precursor gas $Pt(PF_3)_4$ is used to create a $\geq 200$ nm thick Pt dome.

In Fig. 8, we show a cross section of an etched FePt lattice structure with 200 nm lattice spacing. This cross section was made at the position where the trench separation is at its smallest (see also Fig. 6). By inspecting the lattice at various points and measuring the edge slope, here $\alpha = 34(2)°$, we can conclude that the structures stand freely.

### B. SEM images of finished lattice structures

After patterning, the chips were capped by a 50 nm layer of Pt to create a capping layer to provide a uniform electric potential and reflect optical beams in the future experiments. SEM images of all the written lattices were taken before and after this capping layer was deposited. In Fig. 9, we show one of these structures, which contains both large and small lattices, written at the highest resolution. This tapered structure with a lattice spacing that decreases by 1% per line from 5000 nm to 250 nm in 300 lines is the largest structure that was created. The figure shows three different sections of the lattice and an overview of the entire structure (d). The lattice is surrounded by a "background" lattice, which is written with a lower resolution. This background lattice prevents unwanted edge effects to affect the atom traps near the boundaries of the taper. For a complete description of the trap parameters of this tapered lattice, we refer to Ref. 25. On the lower side of the taper, one can trap atoms in the large 5 $\mu$m lattice with a magnetic bias field of approximately 7 G. This traps atoms at 2.5 $\mu$m from the chip with a 5 G trap depth. To trap atoms in the small 250 nm lattice at the other end of the taper, large fields of up to 200 G are required to trap atoms in a 250 nm lattice, 125 nm above the surface of the chip.

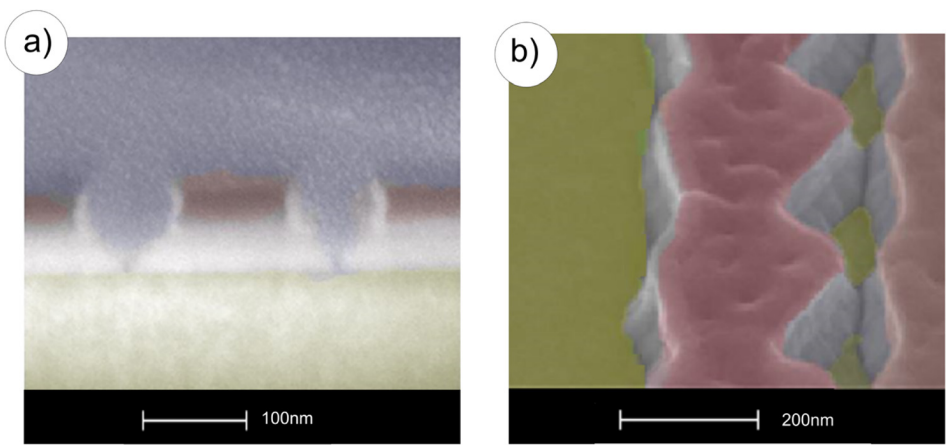

FIG. 8. Inspection of a 200 nm lattice of FePt. (a) FIB cross section; one can distinguish the different materials, from the MgO substrate up: MgO(yellow), FePt (very bright), redeposited triangular shaped metal (also bright, $\approx 100$ nm high), resist (dark red), Pt cover layer (gray). The cross section is taken at a point where the magnet is widest and the trench smallest. b) A SEM image of the same structure before the cross section was made. The scale bars are 100 nm (a) and 200 nm (b).

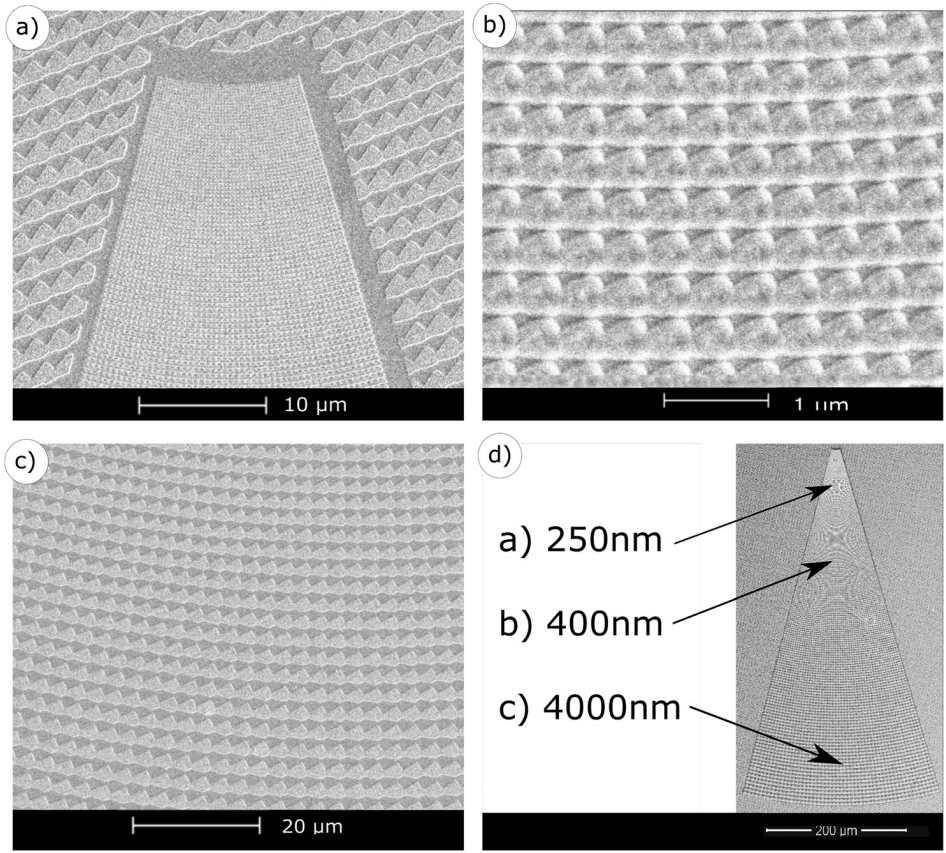

FIG. 9. SEM images of the full magnetic tapered lattice. (a) Tip of the tapered lattice with the surrounding background lattice of 5 $\mu$m. (b) Middle section of the tapered lattice around 400 nm. (c) Lower section of the tapered lattice around 4000 nm. (d) The full tapered lattice ranging from 5 $\mu$m to 250 nm.

In Fig. 10, a Kagome lattice is shown that was written with the same resolution as the taper. In Fig. 10(a) a small section of the unit cell is imaged. Here the resolution of the electron beam can be seen as microscopic dots of approximately 30 nm. In Fig. 10(b), a larger part of the lattice is imaged to show the Kagome lattice and the inter trap lattice spacing of 5 $\mu$m. For a more detailed description of this lattice, we refer to Ref. 12.

## V. SUMMARY

We created lattices in permanent magnetic material to trap ultracold atoms. Molecular beam epitaxy was used to grow 50 nm thick mono-crystalline films of FePt in the $L1_0$ phase with a high remanent magnetization of 900 kA/m. These films were then patterned with e-beam lithography to create the lattices ranging continuously from 5 $\mu$m to 250 nm lattice spacing. A negative tone e-beam resisting with a 30 nm resolution is used as a mask for plasma etching the film. Ar plasma etching was performed to transfer the pattern under optimal conditions into the magnetic film. A Pt cover layer was evaporated, and the resulting structures were studied by taking FIB cross sections. The structures created with this chip will be used in the near future to study the systems of trapped and interacting ultracold Rb atoms.

In the future, the lattices with periods down to 100 nm could be created using thinner magnetic films and thinner resist masks. Below 100 nm, the required resolution is of the order of several nanometer and different methods need to be investigated. Possible techniques are as follows: (1) a multi step process based on a metal mask[21,35] and HSQ as a resist, (2) ion milling with $Ga^+$ as investigated in Ref. 36, (3) ion beam etching (4), or ion milling with a higher resolution $He^+$ beam.

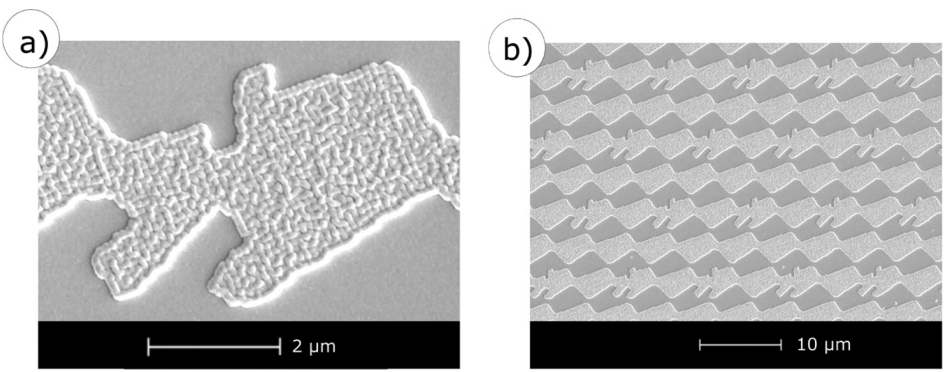

FIG. 10. SEM images of a 5 $\mu$m Kagome lattice. (a) High resolution image of a small section of the unit cell showing the 30 nm resolution with which the lattice is created. (b) Larger section of the lattice showing tens of trap sites that form a Kagome lattice.

## ACKNOWLEDGMENTS

We thank the entire staff of the Kavli Nanolab for their support in the patterning process. We thank Hugo Schlatter and the AMOLF NanoLab Amsterdam for their support in characterizing the magnetic lattices. This work has been supported by the Netherlands Organization for Scientific Research (NWO), the Fund for Scientific Research, Flanders (FWO), and the Hercules Foundation and GOA/14/007 (BOF-KU Leuven). We also acknowledge the support by the EU H2020 FET Proactive project RySQ (640378).